\documentclass[journal,article,submit,pdftex,moreauthors]{Definitions/mdpi} 
\firstpage{1} 
\pubvolume{1}
\issuenum{1}
\articlenumber{0}
\pubyear{2025}
\copyrightyear{2025}
\datereceived{ } 
\daterevised{ } % Comment out if no revised date
\dateaccepted{ } 
\datepublished{ } 
\hreflink{https://doi.org/} % If needed use \linebreak
\preto{\abstractkeywords}{\nolinenumbers}

\Title{Systematic analysis of double Gamow-Teller sum rules}

\TitleCitation{Title}

\Author{Hong-Jin Xie $^{1}$, Yi Lu $^{1}$*\orcidA{}, Shu-Yuan Liang $^{1}$, Yang Lei $^{2}$*\orcidB{} and Calvin W. Johnson $^{3}$*}

\AuthorNames{Hong-Jin Xie, Yi Lu, Shu-Yuan Liang, Yang Lei and Calvin W. Johnson}

\AuthorCitation{Xie, H.; Lu, Y.; Liang, S.; Lei, Y.; Johnson, Calvin W.}

\address{%
$^{1}$ \quad College of Physics and Engineering, Qufu Normal University, 57 Jingxuan West Road, Qufu, Shandong 273165, China\\
$^{2}$ \quad School of National Defense Science and Technology, Southwest University of Science and Technology, Mianyang 621010, China \\
$^{3}$ \quad Department of Physics, San Diego State University, 5500 Campanile Drive, San Diego, CA 92182, USA
}

\corres{Correspondence: luyi@qfnu.edu.cn; leiyang19850228@gmail.com; cjohnson@sdsu.edu}

\abstract{Sum rules are a useful characterization of transition strength functions for atomic nuclei.
	Unlike the Ikeda sum rule for single Gamow-Teller transitions, as a result of SU(4) breaking double Gamow-Teller transition sum rules depend upon the detailed many-body wavefunctions.
	In order to systematically investigate the double Gamow-Teller (DGT) transition sum rules, 
	we approximate the shell model ground state with nucleon-pair condensates, with angular-momentum projection after variation and use expectation values to compute the $2\beta^-$ and $2\beta^+$ sum rules.
	For even-even nuclei in the $1s0d$ and $1p0f$ valence spaces, we quantitatively estimate the model-dependent fractions of the DGT sum rules, and analyze the importance of the double isospin-analogue state to the DGT strength function, relative to SU(4) predictions.}


\begin{document}

%%%%%%%%%%%%%%%%%%%%%%%%%%%%%%%%%%%%%%%%%%

%\setcounter{section}{-1} %% Remove this when starting to work on the template.

\section{Introduction}\label{sec:introduction}

While the  deformation of atomic nuclei has been revealed through electric quadrupole static moments and transitions, 
the spin-isospin content of  nuclei is illuminated by other nuclear properties such as  weak-interaction $\beta$ decays and electromagnetic $M1$  decays and static moments.
%The spin-isospin degree of freedom is an intriguing facet of atomic nuclear structure, related to nuclear $\beta$ decays, $M1$ $\gamma$ transitions and so on.
The single charge-exchange reaction (SCX), e.g. ($p$, $n$) or ($^3$He, $t$), has been a useful probe of Gamow-Teller transition strengths~\cite{Osterfeld1992}, shedding light on the axial charge quenching problem~\cite{ICHIMURA2006446} as well as details of the Gamow-Teller giant resonance~\cite{Fujita2011}.
Double charge exchange reactions via pions ($\pi$DCX) have led to discoveries of new giant resonances, such as dipole resonance built on the isospin analogue state (GDR$\otimes$IAS), and double giant dipole resonance (DGDR)~\cite{Mordechai1991}.
In recent years, improvements in experimental facilities and techniques open new possibilities of heavy-ion double charge-exchange experiments (HIDCX) such as ($^{20}$Ne, $^{20}$O) reactions~\cite{Lenske2019,Cappuzzello2023}.

In recent years neutrinoless double-beta ($0\nu\beta\beta$) decay matrix elements have been sought experimentally and intensely studied theoretically. Calculations have suggested
 empirical correlations between $0\nu\beta\beta$ matrix elements and double Gamow-Teller (DGT), raising the possibility that measurement of DCX cross sections could help reducing the uncertainties of $0\nu\beta\beta$ matrix elements~\cite{Shimizu2018}.
In the shell model, the DGT strengths can be computed state by state~\cite{Auerbach1989,Auerbach2018,RocaMaza2020}, or with the Lanczos strength function technique~\cite{Shimizu2018,Whitehead1980,RevModPhys.77.427,PhysRevC.89.064317}.
%It is notable that the theoretical DGT strength functions appear to have more than one peak, with widths of several MeV \cite{Auerbach2018,Shimizu2018}.

Aside from details of the strength distribution, sum rules provide  important gross characterizations of strength functions, for a given initial state and transition channel. 
For  single Gamow-Teller transitions, i.e. $\hat{O}_{GT\pm} = {\vec{\sigma}}\tau_\pm$, the Ikeda sum rule~\cite{Kiyomi1964} gives the difference between GT- total strength and GT+ total strength in terms of solely the proton and neutron numbers, 
\begin{equation}\label{eqn:Ikeda}
S_{GT-} - S_{GT+} = 3(N-Z).
\end{equation}
This sum rule is useful in validating computer codes and in finding ``missing strength,'' including the quenching of the axial coupling $g_A$,  in experiments.

While the single Gamow-Teller Ikeda sum rule is strictly model-independent, i.e., does  not depend upon the details of the nuclear wave functions, by way of contrast the DGT sum rules have both 
model-independent and model-dependent terms, which we summarize in Sec.~\ref{sec:DGTsumrules}.
In this work, we systematically investigate both $S_{DGT+}$ and $S_{DGT-}$, in a series of even-even nuclei in $1s0d, 1p0f$ major shells.
We quantitatively assess the relative weight of the model-independent terms in the DGT sum rules, as a function of the neutron excess $N-Z$.
We also analyze the DGT strength on the double isospin analogue final state (DIAS) of those nuclei, in comparison with the SU(4) predictions.
%Then we focus on a few candidate nuclei of experimental interest, and predict their GT and DGT sum rules.
 
In Sec.~\ref{sec-GTSR} we present the theoretical framework of projection after variation of nucleon-pair condensates (PVPC), and benchmark one-body spin-isospin transition sum rules of PVPC, against shell-model results. Such a study 
also gauges the robustness of the model-dependent components.
In Sec. \ref{sec-DGTSR} we present systematic analysis of DGT sum rules from PVPC, and analyze the evolution of $S_{DGT\pm}$ as $N-Z$ increases. 
%Then we predict GT/DGT sum rules, and DGT strengths on the double isospin analogue states, for nuclei of experimental interest. 
In Sec. \ref{sec-summary} we draw major conclusions and discuss possibilities of improving the accuracy of our predictions in future work.

\section{Model-independent and model-dependent sum rules}
\label{sec:DGTsumrules}

For the double Gamow-Teller transition, i.e. $\hat{O}^{DGT \pm}_{J\mu} =[\hat{O}^{GT \pm} \otimes \hat{O}^{GT\pm}]_{J\mu}$, sum rules have been derived for both the $J=0$~\cite{1988Sum,Zheng1989} and the $J=2$~\cite{MUTO199213,Sagawa2016} channels.
We confirm that the DGT sum rule formulas in Ref.~\cite{MUTO199213,Sagawa2016} are reproducible while the DGT sum rule formulas in Ref.~\cite{1988Sum,Zheng1989} are 3 times too large.
The DGT sum rules are~\cite{MUTO199213,Sagawa2016}
\begin{eqnarray}
S^{J=0}_{DGT} = S^{J=0}_{DGT-} - S^{J=0}_{DGT+} &=& 2(N-Z)(N-Z+1) + \frac{4}{3}(N-Z)S_{GT+}\nonumber \\
&& - \frac{4}{3} S_\sigma - \frac{4i}{3}\langle \psi_{g.s.} | \vec{\Sigma}\cdot(\hat{O}_{GT-} \times \hat{O}_{GT+}) | \psi_{g.s.} \rangle, \label{eqn:DGTSR-J0} \\
S^{J=2}_{DGT} = S^{J=2}_{DGT-} - S^{J=2}_{DGT+} &=& 10(N-Z)(N-Z-2) + \frac{20}{3}(N-Z)S_{GT+} 
\nonumber \\
&& + \frac{10}{3} S_\sigma + \frac{10i}{3}\langle \psi_{g.s.} | \vec{\Sigma}\cdot(\hat{O}_{GT-} \times \hat{O}_{GT+}) | \psi_{g.s.} \rangle, \label{eqn:DGTSR-J2}
\end{eqnarray}
where $|\psi_{g.s.} \rangle$ is the ground state of the parent nucleus, and $S_{\sigma}$ is the sum rule of a spin transition $\vec{\Sigma} = \sum_k {\vec{\sigma}}(k)$:
\begin{equation}
S_\sigma = \langle \psi_{g.s.} | \vec{\Sigma} \cdot \vec{\Sigma} | \psi_{g.s.} \rangle.
\end{equation}
Note that because the scalar triple product $\vec{\Sigma}\cdot(\hat{O}_{GT-} \times \hat{O}_{GT+})$ is anti-Hermitian, the expectation value $\langle \psi_{g.s.} | \vec{\Sigma}\cdot(\hat{O}_{GT-} \times \hat{O}_{GT+}) | \psi_{g.s.} \rangle$ is imaginary; hence the factor $i$ in in Eq. (\ref{eqn:DGTSR-J0}, \ref{eqn:DGTSR-J2}).

For $N \gg Z$ nuclei, 
%and the term $\langle i | \vec{\Sigma} \cdot \vec{\Sigma} | i \rangle$ is limited, 
%the most substantial part of 
one expects the DGT sum rules to be dominated by the model-independent terms (MIT), determined solely by the proton and neutron numbers $Z$ and $N$.
The remainder depends upon the initial state $|\psi_{g.s.}\rangle$.
%and also the DGT strength to the double-isospin-analogue state (DIAS) \cite{PhysRevC.94.064325}.
The terms involving $\hat{O}_{GT+}|\psi_{g.s.}\rangle$, including $S_{GT+}$, should be suppressed when $N \gg Z$, while $S_\sigma$ measures the spin content. 
Numerical results for even-$A$ Neon isotopes show that when $N-Z\geq 4$ the model-independent component exhausts more than 80\% of the sum rule \cite{MUTO199213}.
Results for semi-magic nuclei, e.g. He, O, Ca isotopes, support this conclusion\cite{Sagawa2016}.

 Wigner's SU(4) symmetry is well known to be broken, but nonetheless provides insights on such transitions.
%If one assumes the SU(4) symmetry is correct, $S_{GT+}=0$ and $S_{GT-}=3(N-Z)$ is concentrated on a single state with $S=1, T_f=(N-Z)/2-1$.
For even-even nuclei with $N-Z \geq 2$, the initial ground state is expected to be dominated by $(S = 0, T = (N-Z)/2)$; in the final $(N-2,Z+2)$ nucleus there are two $S=0$ final states belonging to the same supermultiplet~\cite{1988Sum}, the double-isospin-analogue state (DIAS) and a DGT state with $(S=0,T=(N-Z)/2-2 )$. These latter two yield DGT strengths :
\begin{eqnarray}
B(DGT,  0^+_1 \rightarrow DIAS) = 2(N-Z)\frac{3}{N-Z-1}, \label{eqn:SU4-DGT-DIAS} \\
B(DGT,  0^+_1 \rightarrow DGTS(S=0) ) =  2(N-Z) \frac{ (N-Z)^2 -4 }{N-Z-1}.
\label{eqn:SU4-DGT-DGTS0}
\end{eqnarray}
Summing, we get the model-independent term of Eq. (\ref{eqn:DGTSR-J0}), which also predicts that, when $N-Z=2,4,6,\cdots$, the DGT strength on the DIAS decreases from a maximum of 12, approaching constant $6$.

A model-independent rule can be further derived from Eq.(\ref{eqn:DGTSR-J0}, \ref{eqn:DGTSR-J2}), when $N-Z >0$,
\begin{eqnarray}
S_{GT+} = \frac{ 5 S^{J=0}_{DGT} + 2 S^{J=2}_{DGT} }{20(N-Z)} - \frac{3}{2}(N-Z-1), 
\label{eqn:3GTSR1}
\end{eqnarray}
or with regard to $S_{GT-}$,
\begin{eqnarray}
S_{GT-} = S_{GT+} + 3(N-Z) = \frac{ 5 S^{J=0}_{DGT} + 2 S^{J=2}_{DGT} }{20(N-Z)} + \frac{3}{2}(N-Z+1).
\label{eqn:3GTSR2}
\end{eqnarray}
This model-independent constraint can be used directly by experimentalists. 
If the three sum rules $S_{GT-}, S^{J=0,2}_{DGT}$ can be obtained experimentally, the degree of their agreement with this rule may signal the comparison between one-body quenching in $S_{GT-}$ and two-body quenching in $S^{J=0,2}_{DGT}$.

\section{Formalism and benchmark}\label{sec-GTSR}

We compute the DGT sum rules as expectation values. Our primary model for nuclear wave functions is angular-momentum projected (after variation) nucleon-pair condensate (PVPC) in a shell-model valence space. This is an efficient approximation 
to the full configuration-interaction shell model.

\subsection{Formalism: configuration-interaction shell model}

The underlying conceptual framework for our many-body theory is the spherical shell model. Nucleons can occupy single-particle orbitals with good angular momentum as well as other relevant quantum numbers. 
A flexible instantiation of this framework is the configuration-interaction method~\cite{RevModPhys.77.427}, where one expands a wave function in an orthonormal basis:
\begin{equation}
| \Psi \rangle = \sum_\alpha c_\alpha | \alpha \rangle.
\end{equation}
A convenient choice for the many-body basis states are the occupation representation of Slater determinants.  In the so-called $M$-scheme, these basis states have fixed total $z$-component of angular momentum, or $M$, which is 
easy to enforce. Furthermore, matrix elements of the many-body Hamiltonian, $\langle \alpha | \hat{H} | \beta \rangle$, can be computed very efficiently in the $M$-scheme. 
Here the Hamiltonian can be represented in second quantization, with one-body and two-body terms,
\begin{equation}\label{eqn:Hamiltonian}
\hat{H} = \sum_a \epsilon_a \hat{n}_a + \sum_{abcd} \frac{ \sqrt{1+\delta_{ab}} \sqrt{ 1+ \delta_{cd} } }{4} \sum_{IT} V(abcd; I T) \sum_{M M_T} \hat{A}^\dagger_{IM, TM_T} (ab) \hat{A}_{IM, T M_T}(cd), 
\end{equation}
where $\hat{A}^\dagger \equiv \sum_{\alpha\beta} A_{\alpha\beta} \hat{c}^\dagger_\alpha \hat{c}^\dagger_\beta$, and $\alpha, \beta$ stand for (usually harmonic oscillator) orbitals with quantum numbers e.g. $(n_\alpha l_\alpha j_\alpha m_\alpha)$.
In the standard configuration-interaction shell model, the coefficients $A_{\alpha\beta}$ are simply Clebsch-Gordan coefficients, but below, in the nucleon-pair condensate model, they are determined variationally.
We use the  code {BIGSTICK}~\cite{2013Factorization} for full configuration-interaction calculations or otherwise take results from the literature.

For this work, we work in three model spaces, each with its own shell model effective interaction, stored as single-particle energies and two-body matrix elements.
In the $0p$ shell, assuming a frozen $^4$He core, we use 
the CK\uppercase\expandafter{\romannumeral 2} interaction~\cite{Cohen1965}; in the $1s0d$ shell, on top of  a frozen $^{16}$O core, we use USDB~\cite{PhysRevC.74.034315}; and 
finally in the  $1p0f$ shell, we use the interaction GX1A~\cite{Honma2005}.

Although the basis states have good $M$ and not individually good total angular momentum $J$, because the Hamiltonian is rotationally invariant, eigenstates with good $J$ naturally arise. 
The downside of the $M$-scheme is that one need large dimensions to 
build in the necessary correlations. 

\subsection{Formalism: projection after variation of a nucleon-pair condensates}

Because of the demands of large basis dimension for full-configuration shell model calculations, we also use an efficient alternative.
For an even-even nucleus with $N_\pi$ valence protons and $N_\nu$ valence neutrons, a general nucleon-pair condensate (PC) is defined as
\begin{equation}
|PC\rangle \equiv \frac{1}{(N_\pi/2)! (N_\nu/2)!} ( \hat{A}^\dagger_{\pi} )^{N_\pi /2} (\hat{A}^\dagger_{\nu} )^{N_\nu /2} | 0 \rangle,
\end{equation}
%where $\hat{A}^\dagger \equiv \sum_{\alpha\beta} A_{\alpha\beta} \hat{c}^\dagger_\alpha \hat{c}^\dagger_\beta$, $\alpha, \beta$ stand for harmonic oscillator orbits with quantum numbers e.g. $(n_\alpha l_\alpha j_\alpha m_\alpha)$, 
Here $A_{\alpha\beta}$ is the ``pair structure coefficients", to be determined variationally.
%Given shell-model effective interactions, with one-body and two-body terms,
%\begin{equation}\label{eqn:Hamiltonian}
%\hat{H} = \sum_a \epsilon_a \hat{n}_a + \sum_{abcd} \frac{ \sqrt{1+\delta_{ab}} \sqrt{ 1+ \delta_{cd} } }{4} \sum_{IT} V(abcd; I T) \sum_{M M_T} \hat{A}_{IM, TM_T} (ab) \hat{A}_{IM, T M_T}(cd),
%\end{equation}
Given the shell-model effective Hamiltonian, Eq.~(\ref{eqn:Hamiltonian}), 
the trial wavefunction $|PC\rangle$ is optimized by varying the pair structure coefficients $A^\pi_{\alpha\beta}$, $A^\nu_{\alpha\beta}$, so that the expectation value of Hamiltonian is minimized~\cite{mcsm-pair,PhysRevC.44.R598,PhysRevC.26.2640,PhysRevC.102.024310,PhysRevC.105.034317},
\begin{equation}
\min_{A^\pi_{\alpha\beta}, A^\nu_{\alpha\beta}} \langle PC | \hat{H} | PC \rangle / \langle PC | PC \rangle.
\end{equation}
For example, in the $1s0d$ major shell, there are $C^{12}_2 = 66$ free parameters for $A^\pi_{\alpha\beta}$ and $A^\nu_{\alpha\beta}$ respectively.
With the formulas derived in Ref.~\cite{PhysRevC.105.034317} and the multi-variable minimizers in the GNU Scientific Library codes~\cite{gsl}, such a variation converges after typically about 200 iteractions.

The Hamiltonian in Eq.~(\ref{eqn:Hamiltonian}) respects angular momenta as good quantum numbers, corresponding to rotational symmetry, but in the variational process, the rotational symmetry is spontaneously broken, and the pair condensate ends up without good angular momenta.
Therefore  angular momentum projection is implemented to restore rotational symmetry. %, before comparison with experimental data.
Projected bases with angular momentum $(J,M)$ can be constructed from the variationally optimized pair condensate (VPC),
\begin{equation}
\hat{P}^J_{MK} |  VPC
\rangle,~~~  K = -J, \cdots, J,
\end{equation}
using the angular momentum projection operator $\hat{P}^J_{MK} = \frac{2J+1}{8\pi}\int D^{J*}_{MK}(\Omega)\hat{R}(\Omega) d\Omega$~\cite{PhysRevC.105.034317}. Here $D^{J}_{MK}(\Omega)$ is the Wigner D-function and  the rotation operator is $\hat{R}(\Omega)=\hat{R}(\alpha,\beta,\gamma)= \exp(i\gamma \hat{J}_z) \exp(i\beta \hat{J}_y) \exp(i\alpha \hat{J}_z)$.
Nuclear states are approximated as linear combinations of such bases, and the mixing among such bases are introduced by solving the so-called Hill-Wheeler equation, a generalized eigenvalue problem,
\begin{equation}\label{eqn-HillWheeler}
\sum_K \mathcal{H}^J_{K' K} g^r_{JK} = \epsilon_{r,J} \sum_K \mathcal{N}^J_{K'K} g^r_{JK},
\end{equation}
where $\mathcal{H}^J_{K' K} \equiv \langle PC | \hat{H}\hat{P}^J_{K'K} |PC \rangle$ and $\mathcal{N}^J_{K' K} \equiv \langle PC | \hat{P}^J_{K'K} | PC \rangle$ are the Hamiltonian matrix elements and overlaps of projected bases respectively, $\epsilon_{r,J}$ is the eigen-energy of the r-th state with angular momentum $J$, with at most $(2J+1)$ states, and $g^r_{JK}$ determines the eigen-wavefunctions,
\begin{equation} \label{eqn:projected-wave}
\psi^r_{JM} = \sum_K g^r_{JK} \hat{P}^J_{MK}|PC\rangle.
\end{equation}
Throughout this work, we are restricted to one major shell, without considerations about cross-shell configurations.
The PVPC ground state could be considered as close to an HFB vacuum with conserved particle numbers, see the discussions on the canonical transformation of a pair condensate in \cite{Liang2025}, without including quasi-particle excitations.

\subsection{Formalism: sum rules as expectation values}

\label{sec:sumrules}

An advantage of certain sum rules, such as the non-energy-weighted sum rules investigated here, is that they can be rewritten as expectation values of operators. This is particularly straightforward for 
sum rules of one-body operators.
Given an initial state $|J_i M_i \rangle$ and a one-body transition operator $\hat{O}_{t\tau}$, the sum rule is summed over $J_f, M_f$ of the final state $|J_f M_f \rangle$, and averaged over $M_i$,
\begin{equation}
S = \frac{1}{2J_i + 1} \sum_{J_f M_f \tau M_i} |\langle J_f M_f | \hat{O}_{t \tau} | J_i M_i \rangle |^2 = \langle J_i M_i |(-1)^t [t] (\hat{O}_t \otimes \hat{O}_t)_{00} | J_i M_i \rangle.
\end{equation}
In other words, the sum rule is the expectation value of an sum rule operator
\begin{equation}
\hat{O}_{SR} \equiv (-1)^t [t] (\hat{O}_t \otimes \hat{O}_t)_{00}.
\end{equation}
Such an operator can be written into a Hamiltonian-like form in the occupation space, and fed to structure codes for fast evaluation of sum rules~\cite{Johnson2015Systematics, PhysRevC.97.034330},
\begin{equation}
\hat{O}_{SR} = \sum_{ab} g_{ab} [j_a] (a^\dagger \otimes \tilde{b})_{00} + \frac{1}{4} \sum_{abcdJ} \zeta_{ab}\zeta_{cd} W_J(abcd)[J] (\hat{A}^\dagger_J(ab) \otimes \tilde{A}_J(cd))_{00},  \label{eq:OSR}
\end{equation}
where $[j_a] \equiv \sqrt{2j_a+1}$, $\zeta_{ab} \equiv \sqrt{1+\delta_{ab} }$, $\hat{A}_J(ab) \equiv (\hat{a}^\dagger \otimes \hat{b}^\dagger)_{JM}$, $a,b$ index single-j orbits with quantum numbers $(nlj)$.
Given the transition operator $\hat{O}_t$, the values of $g_{ab}$ and $W_J(abcd)$ can be computed with the code {\it PandasCommute} which is publicly available~\cite{PhysRevC.97.034330}.

For transitions involving operators of rank higher than one-body, e.g., DGT, the basic idea still holds but the details are more complicated. We address this in Sec.~\ref{sec:DGTsumrule}.

\subsection{Benchmark of $S_{GT+}$ and $S_\sigma$}

\label{sec:benchmark}

The approximate ground state from PVPC has an energy above that of the ``exact" shell model ground state, by typically $0.5$ MeV for semi-magic nuclei, and $1\sim 2$ MeV for open-shell nuclei~\cite{PhysRevC.105.034317}.
Shell model sum rules show secular dependence on the excitation energy of the initial state, although different configurations can lead to local fluctuations of the sum rules~\cite{Johnson2015Systematics}.
As a result, approximate ground states, which can be considered as a linear combination of low-energy shell-model configurations, end up with reasonable sum rules for E2, M1 and Gamow-Teller transitions~\cite{Johnson_2020}.
Nonetheless, spin-isospin transitions are well known to be sensitive to details of the many-body wavefunction, compared with electromagnetic transitions, hence we benchmark $S_{GT+}$ and $S_\sigma$ as in Eq.~(\ref{eqn:DGTSR-J0})(\ref{eqn:DGTSR-J2}), before analyzing $S_{DGT\pm}$.

The expectation value of $\hat{O}_{SR}$ from Eq.~(\ref{eq:OSR}) for the projected wavefunction  in Eq.~(\ref{eqn:projected-wave}) can be written
\begin{equation} \label{eqn:SR-projected-wave}
\langle \psi^r_{J_i M_i} | \hat{O}_{SR} | \psi^r_{J_i M_i} \rangle = \sum_{ K' K} g^r_{J_i K'} g^r_{J_i K} \langle PC | \hat{O}_{SR} \hat{P}^{J_i}_{K'K} | PC \rangle,
\end{equation}
where $\hat{P}^{J_i}_{K'M_i} \hat{O}_{SR} \hat{P}^{J_i}_{M_i K} = \hat{O}_{SR} \hat{P}^{J_i}_{K'K}$ is used, because $\hat{O}_{SR}$ is a scalar and $\hat{P}^{J_i}_{ K' M_i} \hat{P}^{J_i}_{M_i K} = \hat{P}^{J_i}_{K' K}$.
Expressions such as $\langle PC | \hat{O}_{SR} \hat{P}^{J_i}_{K'K} | PC \rangle$ are in the form of Hamiltonian ``kernel" values $\mathcal{H}^J_{K'K}$ in (\ref{eqn-HillWheeler}), therefore can be easily computed.

\begin{figure}[ht!] 
	\centering\includegraphics[width=0.8\textwidth]{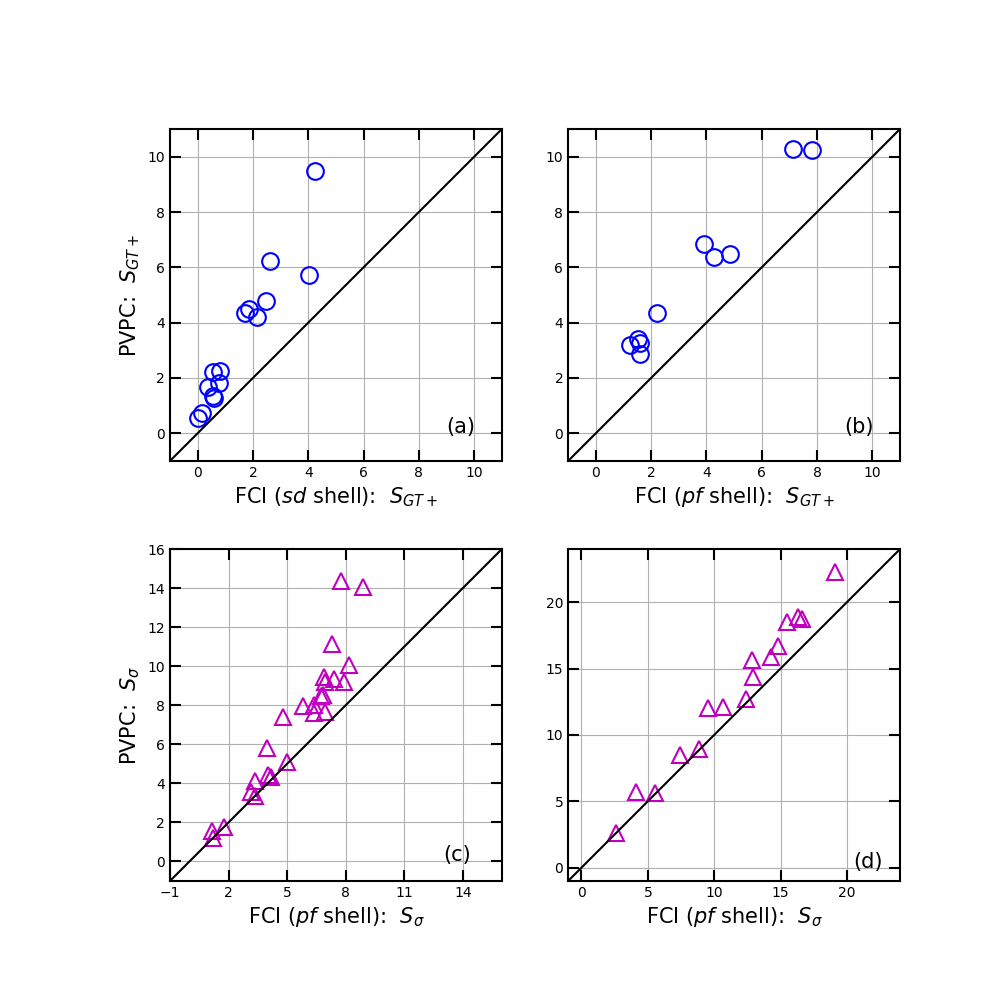}
	\caption{ Benchmark of one-body spin-isospin sum rules of ground states generated from PVPC, against that of shell-model ground states by ``exact" diagonalization. ``PVPC" stands for ``projection after variation of a pair condensate". Panel (a): GT+ sum rules $S_{GT+}$ of even-even nuclei in the $1s0d$ major shell; Panel (b): $S_{GT+}$ of even-even nuclei in the $1p0f$ major shell; Panel (c): $S_\sigma$ of even-even nuclei in the $1s0d$ major shell; Panel (d): $S_\sigma$ of even-even nuclei in the $1p0f$ major shell.}
	\label{fig:SGT+Ssigma}
\end{figure}

In Fig. \ref{fig:SGT+Ssigma} we show numerical results of $S_{GT+}$ and $S_{\sigma}$ for even-even nuclei in the $1s0d$ and $1p0f$ major shells.
The horizontal values are shell-model sum rules generated with the BIGSTICK code~\cite{2013Factorization}, while the vertical values are sum rules generated with the PVPC model.
The diagonal solid line means the approximate sum rules are perfectly in agreement with shell-model results.
We see that $S_{GT+}$ is over-estimated by PVPC, as in other approximate methods such as the Projected Hartree-Fock method and the nucleon-pair approximation~\cite{Johnsonb}.
In the $1s0d$ major shell, the PVPC $S_{GT+}$ can be twice that of shell-model $S_{GT+}$, while in the $1p0f$ major shell, the PVPC $S_{GT+}$ is about 1.2$\sim$1.5 times that of the shell model.
Previous shell-model investigations~\cite{Johnson2015Systematics} show that, in the $1s0d$ shell the Gamow-Teller sum rule typically increases as the excitation of the initial energy increases. 
As our approximate ground state may include excited configurations other than the pure shell-model ground state, the overestimation of $S_{GT+}$ agrees with that trend.

However $S_{\sigma}$ appears less sensitive, as PVPC $S_\sigma$ values are remarkably faithful to shell-model $S_\sigma$ values. 
We speculate this is because the transition $\hat{\bf\Sigma} = \sum_k \vec{\sigma}(k)$ involves only spin components, not spin-isospin components as the Gamow-Teller transition does.

\section{Double Gamow-Teller sum rules}\label{sec-DGTSR}
\subsection{Formalism}

\label{sec:DGTsumrule}

While Ref.~\cite{1988Sum,Zheng1989,MUTO199213,Sagawa2016} study the difference between $S_{DGT-}$ and $S_{DGT+}$, in this work we compute $S_{DGT\pm}$ directly.
Invoking closure condition $\sum_{J_f M_f} |J_f M_f \rangle \langle J_f M_f | = 1$, $S_{DGT\pm}$ is the expectation value of a four-body operator,
\begin{equation}
S_{DGT\pm} = \sum\limits_{J_f M_f \mu } \mid\langle J_f M_f \mid\hat{O}^{DGT\pm}_{J \mu} \mid J_i M_i \rangle \mid^2
=\langle J_i M_i \mid \sum_\mu  (\hat{O}^{DGT\pm}_{J\mu})^\dagger\hat{O}^{DGT\pm}_{J\mu}\mid J_i M_i \rangle,
\label{eqn:SDGT}
\end{equation}
where $\hat{O}^{DGT\pm}_{J\mu}$ is the double Gamow-Teller transition operator,
\begin{equation}
\hat{O}^{DGT \pm}_{J\mu} =[\hat{O}^{GT \pm} \otimes \hat{O}^{GT\pm}]_{J\mu},
\end{equation}
$\hat{O}^{GT\pm}$ denotes the Gamow-Teller transition operator, i.e. $\sigma \tau_\pm$, ``$\otimes$" signals angular momentum coupling,
$J = 0, 2$, and $\mu = -J, \cdots, J$.
Note that $J=1$ is forbidden because of symmetry consideration~\cite{Sagawa2016}.

Because of the the four-body nature of the operator, computing the DGT sum rule expectation value is more complicated than for single GT sum rules. We are aided, however, by being able to separate out the proton and neutron components of the operators, 
which makes this calculation tractable.
The DGT transition operator can be expanded as
\begin{equation} \label{eqn:ODGT}
\hat{O}^{DGT \pm}_{J\mu} =[\hat{O}^{GT \pm} \times \hat{O}^{GT\pm}]_{J\mu}
=\sum\limits_{\alpha,\gamma \in \nu/\pi; ~\beta, \delta \in \pi/\nu}C^{DGT}_{J\mu}(\alpha\beta\gamma\delta) \hat{c}^\dagger_\alpha \hat{c}_\beta \hat{c}_\gamma^\dagger \hat{c}_\delta,
%& =-\sum\limits_{i,j,k,l}C_{i,j,k.l} \hat{n}_i^\dagger \hat{n}_k^\dagger \hat{p}_j \hat{p}_l\\
%&=-\sum\limits_{i<k,j<l}[C_{ijkl}-C_{kjil}-C_{ilkj}+C_{klij}] \hat{n}_i^\dagger \hat{n}_k^\dagger \hat{p}_j \hat{p}_l\\
\end{equation}
where $C^{DGT}_{J\mu}(\alpha\beta\gamma\delta)$ is found to be
\begin{equation}
C^{DGT}_{J\mu}(\alpha\beta\gamma\delta) 
= \sum\limits_{M_1} (1,M_1;1,\mu-M_1|J\mu) g^{GT}_{M_1}(\alpha \beta) g^{GT}_{\mu-M_1}(\gamma \delta).
\end{equation}
Here $g^{GT}_M(\alpha\beta)$ is defined such that $\hat{O}^{GT\pm}_{1M} 
=\sum\limits_{\alpha \in \nu/\pi, \beta \in \pi/\nu} g^{GT}_M (\alpha\beta) \hat{c}_{\alpha}^{\dagger} \hat{c}_{\beta}$,
\begin{equation}
g^{GT}_M(\alpha\beta) \equiv (-1)^{j_\beta -m_\beta}\sqrt{\frac{2j_\alpha+1}{3}} \langle \alpha ||\sigma|| \beta \rangle (j_\alpha, m_\alpha; j_\beta, -m_\beta | 1 M ),
\end{equation}
and 
$(j_\alpha, m_\alpha; j_\beta, -m_\beta | 1 M )$ denotes the Clebsch-Gorden coefficients.
We denote single-$j$ orbits by Latin letters $a,b,\cdots,$ with quantum numbers $(n_a l_a j_a)$, $m$-scheme orbits by Greek letters $\alpha,\beta,\cdots,$ with quantum numbers $(n_\alpha l_\alpha j_\alpha m_\alpha)$.

In order to cast the DGT transition operator into computation with pair condensates, we write it in 
%We denote sum rule operators $\hat{O}^{DGTSR\pm}_{J\mu} \equiv \sum_\mu  (\hat{O}^{DGT\pm}_{J\mu})^\dagger\hat{O}^{DGT\pm}_{J\mu}$.
terms of pair creators and pair annihilators, e.g. for DGT-,
\begin{equation}
\hat{O}^{DGT-}_{J\mu} = -\sum\limits_{\alpha<\beta}(\hat{P}^{\alpha\beta}_{\pi})^{\dagger}(\hat{P}^{\alpha\beta}_{\nu}),
\end{equation}
i.e., the operator annihilates a pair of neutrons and creates a pair of protons.
The pair structure coefficients of $(\hat{P}^{\alpha\beta}_{\pi})^{\dagger}$ and $\hat{P}^{\alpha\beta}_{\nu}$ can be derived, 
\begin{eqnarray}
(p^{\alpha\beta}_{\pi})_{\gamma\delta} &=& \delta_{\alpha\gamma} \delta_{\beta\delta} - \delta_{\alpha\delta} \delta_{\beta\gamma}, \\
(p^{\alpha\beta}_{\nu})_{\gamma\delta} &=& C^{DGT}_{J\mu}(\alpha\gamma\beta\delta) - C^{DGT}_{J\mu}(\beta\gamma\alpha\delta) - C^{DGT}_{J\mu}(\alpha\delta\beta\gamma) + C^{DGT}_{J\mu}(\beta\delta\alpha\gamma).
\end{eqnarray}
The sum rule operator for DGT- channel can be also written in terms of pair creators and pair annihilators,
\begin{equation} \label{eqn:ODGTSR}
\hat{O}^{DGTSR-}_{J\mu} = \sum\limits_{\alpha < \beta}(\hat{P}^{\alpha\beta}_{\nu})^{\dagger}(\hat{P}^{\alpha\beta}_{\pi})   \sum\limits_{\gamma < \delta}(\hat{P}^{\gamma\delta}_{\pi})^{\dagger}(\hat{P}^{\gamma\delta}_{\nu}).
\end{equation}
The expectation value of $\hat{O}^{DGTSR-}_{J\mu}$ on the wavefunction Eq.~(\ref{eqn:projected-wave}) is similar to Eq.~(\ref{eqn:SR-projected-wave}),
\begin{eqnarray}
S_{DGT-}(\psi^r_{J_i M_i}) &=& \sum_{K'K} g^r_{J_i K'} g^r_{J_i K} \langle PC 
\mid 
\hat{O}^{DGTSR-}_{J\mu}
\hat{P}^{J_i}_{K'K} \mid PC \rangle \nonumber \\
&=& \sum_{K'K} g^r_{J_i K'} g^r_{J_i K} \frac{8J+1}{8\pi} \int D^{J_i *}_{K'K}(\Omega)
\sum\limits_{m<n,i<k}
\langle \hat{A}_{\pi}^{N_\pi/2}
\mid
(\hat{P}^{mn}_{\pi})(\hat{P}^{ik}_{\pi})^\dagger
\mid
(\hat{A'}_{\pi}^{\dagger})^{N_\pi/2}\rangle \nonumber \\
&&
\times \langle\hat{A}_{\nu}^{N_\nu/2}
\mid
(\hat{P}^{mn}_{\nu})^{\dagger}(\hat{P}^{ik}_{\nu})
\mid
(\hat{A'}_{\nu}^{\dagger})^{N_\nu/2}\rangle d\Omega, 
\label{eqn-SDGT-PVPC}
\end{eqnarray}
where $\hat{A'}_{\pi}^{\dagger} \equiv \hat{R}(\Omega) \hat{A}_{\pi}^{\dagger}$ and 
$\hat{A'}_{\nu}^{\dagger} \equiv \hat{R}(\Omega) \hat{A}_{\nu}^{\dagger}$ are the rotated pairs.
Terms such as $\langle \hat{A}_{\pi}^{N_\pi/2}
\mid
(\hat{P}^{mn}_{\pi})(\hat{P}^{ik}_{\pi})^\dagger
\mid
(\hat{A'}_{\pi}^{\dagger})^{N_\pi/2}\rangle$ and $\langle\hat{A}_{\nu}^{N_\nu/2}
\mid
(\hat{P}^{mn}_{\nu})^{\dagger}(\hat{P}^{ik}_{\nu})
\mid
(\hat{A'}_{\nu}^{\dagger})^{N_\nu/2}\rangle$ can be computed with overlap formulas between pair condensates with ``impurity" pairs~\cite{PhysRevC.105.034317}.

\subsection{Benchmark DGT sum rules}

We benchmark DGT sum rules from PVPC against that from shell-model results available in the literature.
In Ref. \cite{Sagawa2016} shell model DGT sum rules are presented for a series of semi-magic nuclei.
With no valence protons or full valence neutrons in semi-magic nuclei, $S_{DGT+} = 0$, and $\hat{O}_{GT+}|\psi_{g.s.}\rangle = 0$,
therefore Eq.~(\ref{eqn:DGTSR-J0})(\ref{eqn:DGTSR-J2}) can be utilized in the shell model to acquire $S_{DGT-}$, surrounding the difficulty of three-body terms \cite{Sagawa2016}.

In this work, the PVPC $S_{DGT-}$ values are computed with Eq.~(\ref{eqn-SDGT-PVPC}), using the four-body sum rule operator directly.
In Tab.~\ref{tab-DGTSR-benchmark}, we present $S_{DGT-}$ for the semi-magic nuclei studied in Ref.~\cite{Sagawa2016}, from both PVPC and the shell model in comparison.
%The CK\uppercase\expandafter{\romannumeral 2} interactions~\cite{Cohen1965} are used in the $0p$ shell, USDB~\cite{PhysRevC.74.034315} in the $1s0d$ shell, and GX1A~\cite{Honma2005} in the $1p0f$ shell.
It is shown that $S_{DGT-}$ from PVPC is remarkably close to that from the shell model (in the parentheses), which is unsurprising, as it is known that the ground state of a semi-magic nucleus is typically well approximated with a pair condensate.

For open-shell nuclei, shell-model DGT sum rules have not been computed systematically, as far as we know.
Therefore, in the next subsection, we present the PVPC results, and also analyze the error induced by the overestimated $S_{GT+}$, as mentioned previously.
\begin{table} [H]
	\centering
	\caption{ DGT- sum rules generated with the PVPC method, in comparison against shell model results in the parentheses. The shell-model results are quoted from Ref.~\cite{Sagawa2016}. }
	\begin{tabular}{ccccccc}
		\hline\hline
		Initial state		& J=0  	& J=2	 \\
		\hline
		$^{6}He$	& 12.00(12)	& 0.013(0) \\
		$^{8}He$	& 39.54(39.7) & 81.39(80.7)\\
		$^{14}C$	& 7.57(8.98)	& 10.87(7.55) \\
		$^{18}O$	& 10.42(10.4)	& 3.99(3.96) \\
		$^{20}O$	& 33.77(35.5)	& 91.2(91.3) \\
		$^{42}Ca$	& 8.50(8.5)	& 8.75(8.75)\\
		$^{44}Ca$	& 31.68(32.6)	& 100.8(98.5)\\
		$^{46}Ca$	& 70.38(72.3)	& 274.0(269.3)\\
		$^{48}Ca$	& 125.75(135.5)& 525.6(501.2)\\
		\hline\hline
	\end{tabular}

	\label{tab-DGTSR-benchmark}
\end{table}

\subsection{Systematic analysis of $S_{DGT\pm}$} \label{SDGT}
\begin{figure}[ht!]
	\centering\includegraphics[width=1\textwidth]{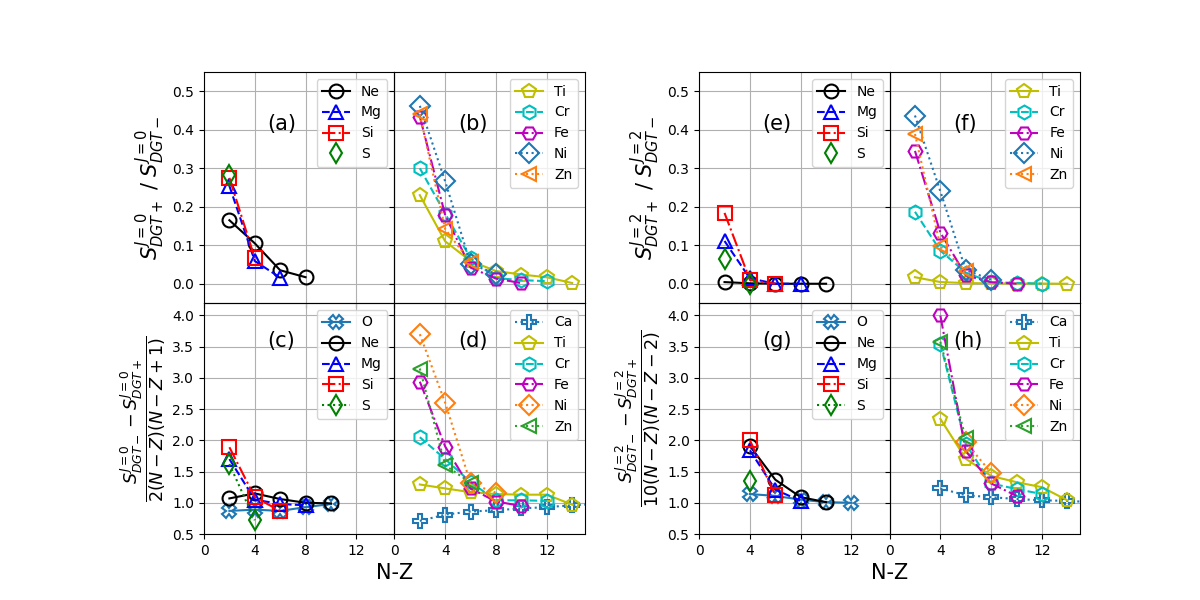}
	\caption{ DGT sum rules of even-even nuclei in the $1s0d$ major shell, from PVPC. Panel (ab): ratio between $S^{J=0}_{DGT+}$ and $S^{J=0}_{DGT-}$; Panel (cd): ratio between $S^{J=0}_{DGT-} - S^{J=0}_{DGT+}$ and the model independent term $2(N-Z)(N-Z+1)$, in Eq.~(\ref{eqn:DGTSR-J0}); Panel (ef): ratio between $S^{J=2}_{DGT+}$ and $S^{J=2}_{DGT-}$;  Panel (gh): ratio between $S^{J=2}_{DGT-} - S^{J=2}_{DGT+}$ and the model independent term $10(N-Z)(N-Z-2)$ in Eq.~(\ref{eqn:DGTSR-J2}).}
	\label{fig:DGTSR}
\end{figure}

We compute $S_{DGT\pm}$ for $N \geq Z$ even-even isotopes of O, Ne, Mg, Si, S, Ca, Ti, Cr, Fe, Ni, Zn in the $1s0d$ and $1p0f$ major shells.
In Fig. \ref{fig:DGTSR} (ab,ef) we show the ratio between $S_{DGT+}$ and $S_{DGT-}$, versus the neutron excess.
When $N-Z=0$ it is trivial, as $S_{DGT+}$ equals $S_{DGT-}$, therefore in these two panels $N-Z$ starts from $2$.
While it is anticipated that $S_{DGT+}$ approaches $0$ as $N-Z$ increases, we show the speed of this tendency.
When $N-Z = 6$, $S_{DGT+}$ is below 7\% of $S_{DGT-}$; when $N-Z=8$, $S_{DGT+}$ is below 4\% of $S_{DGT-}$; in such cases $S_{DGT+}$ is negligible.

In Fig.~\ref{fig:DGTSR} (cd,gh), we show the difference between $S_{DGT-}$ and $S_{DGT+}$, i.e. the ``sum rules" in Ref.~\cite{1988Sum,Zheng1989,MUTO199213,Sagawa2016}, $S_{DGT} = S_{DGT-} - S_{DGT+}$, in comparison with the model independent terms (MIT) in (\ref{eqn:DGTSR-J0})(\ref{eqn:DGTSR-J2}).
Indeed for Neon isotopes, the MIT is faithful to the sum rule in the $J=0$ channel, in consistence with Ref.~\cite{MUTO199213}, but for other nuclei the DGT sum rules show more deviations from the MIT.

In the $J=0$ channel, when $N-Z=2,4$, $S_{DGT-} - S_{DGT+}$ can be $0.7\sim 4$ times the MIT; when $N-Z = 6$, $S_{DGT-} - S_{DGT+}$ is about $0.8\sim 1.5$ times the MIT.
The good news is that when $N-Z \geq 8$ the MIT approaches the DGT sum rule .

Because the MIT term in $S^{J=2}_{DGT}$, $10(N-Z)(N-Z-2)$, equals zero when $N-Z=2$, Panel (gh) starts from $N-Z=4$.
It is noteworthy that in the $J=0$ channel, MIT can be larger or smaller than $S_{DGT}$, while in the $J=2$ channel, MIT is smaller than $S_{DGT}$, and MIT approaches the sum rule faster in the $J=0$ channel.
Also, $1p0f$-shell nuclei have larger DGTSR, compared with $1s0d$-shell nuclei.

In Eq.~(\ref{eqn:DGTSR-J0}, \ref{eqn:DGTSR-J2}), there are three model-dependent contributions: $S_{GT+}, S_\sigma$, and a three-body term.
In our benchmark results, PVPC $S_\sigma$ agrees well with shell-model $S_\sigma$ values, but PVPC overestimates $S_{GT+}$ by about 2 times in the $1s0d$ shell and about 1.2$\sim$1.5 times in the $1p0f$ shell.
The contribution from $S_{GT+}$ scales with $N-Z$, 
i.e. $ \frac{4}{3}(N-Z)S_{GT+}$ for $S^{J=0}_{DGT}$, and $\frac{20}{3}(N-Z)S_{GT+}$ for $S^{J=2}_{DGT}$.
Therefore, the error brought by PVPC $S_{GT+}$ may become significant, when $N-Z$ increases.
We make use of the shell-model $S_{GT+}$ values and introduce corrections,
\begin{eqnarray}
S'^{J=0}_{DGT} = S^{J=0, PVPC}_{DGT} - \frac{4}{3}(N-Z)(S^{PVPC}_{GT+} - S^{FCI}_{GT+}),
\label{eqn:SDGTJ0correc}
 \\
S'^{J=2}_{DGT} = S^{J=2, PVPC}_{DGT} - \frac{20}{3}(N-Z)(S^{PVPC}_{GT+} - S^{FCI}_{GT+}),
\label{eqn:SDGTJ2correc}
\end{eqnarray}
where $S^{PVPC}_{DGT}$ is the computational value from Eq.~(\ref{eqn-SDGT-PVPC}), $S^{PVPC}_{GT+}$ is the $S_{GT+}$ value from PVPC, and $S^{FCI}_{GT+}$ is that from the full configuration-interaction shell model.
In other words, we replace the PVPC $S_{GT+}$ values with shell-model ones, when available, in the DGTSR, as corrections.
Thus the contributions from both $S_{\sigma}$, $S_{GT+}$ are supposed to be reliable, while the contributions from $\langle \psi_{g.s.} | \vec{\Sigma}\cdot(\hat{O}_{GT-} \times \hat{O}_{GT+}) | \psi_{g.s.} \rangle$ are from PVPC.
We denote the corrected DGT sum rules as $S'_{DGT}$ instead of $S_{DGT}$.

\begin{figure}[ht!]
	\centering\includegraphics[width=1\textwidth]{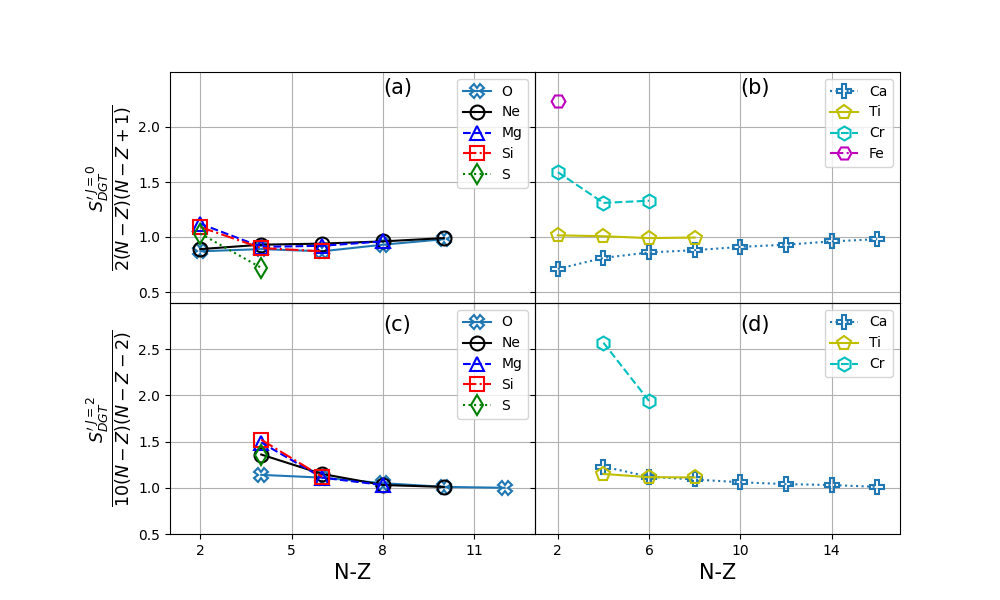}
	\caption{ The ratio between the DGT sum rule and the MIT, with corrections introduced via (\ref{eqn:SDGTJ0correc}, \ref{eqn:SDGTJ2correc}).}
	\label{fig:DGTSRcorr}
\end{figure}

In Fig. \ref{fig:DGTSRcorr}, we present corrected sum rules, versus the MIT term, when shell-model $S_{GT+}$ values are available.
Compared with Fig. \ref{fig:DGTSR}(c-h), the MIT approaches the DGTSR faster (as $N-Z$ increases) after corrections.

\subsection{DGT strength on the DIAS final state}
Among all possible final states, the double isospin-analogue state (DIAS) of the initial state is special.
Assuming the initial state $|J_i M_i\rangle$ has isospin quantum numbers $\left(t = \frac{N-Z}{2}, t_z = \frac{N-Z}{2}\right)$, the DIAS final state is written as,
\begin{equation}
|{\rm DIAS} \rangle = \frac{\hat{T}_- \hat{T}_- |J_i M_i \rangle}{\sqrt{2(N-Z)(N-Z+1)}} ,
\end{equation}
and the DGT- strength from $|J_i M_i \rangle $ to $|{\rm DIAS} \rangle$ is
\begin{equation}\label{eqn:DGT-DIAS}
D_-( DIAS) =  \frac{|\langle J_i M_i | \hat{T}_+ \hat{T}_+ \hat{O}^{J \mu}_{DGT-} | J_i M_i \rangle |^2}{2(N-Z)(N-Z+1)}.
\end{equation}
When there are no valence protons or full valence neutrons in the configuration space, $\hat{T}_+ | J_i M_i \rangle = 0$, by derivations with commutators, it is shown that $D_-( DIAS)$ can be expressed with $S_\sigma, S_{GT\pm}$~\cite{Sagawa2016},
\begin{equation}
D_-(DIAS) = \frac{2 [ S_\sigma - S_{GT-} - S_{GT+}]^2 }{3(N-Z)(N-Z-1)} .
\end{equation}
In general, for open-shell nuclei, following Eq.~(\ref{eqn:DGT-DIAS}), the strength on the DIAS can be computed in terms of expectation value of the four-body operator $\hat{T}_+ \hat{T}_+ \hat{O}^{J \mu}_{DGT-}$.
Therefore it can be computed in a similar way as in Eq.~(\ref{eqn:ODGTSR}).

\begin{figure}[ht!]
	\centering\includegraphics[width=1\textwidth]{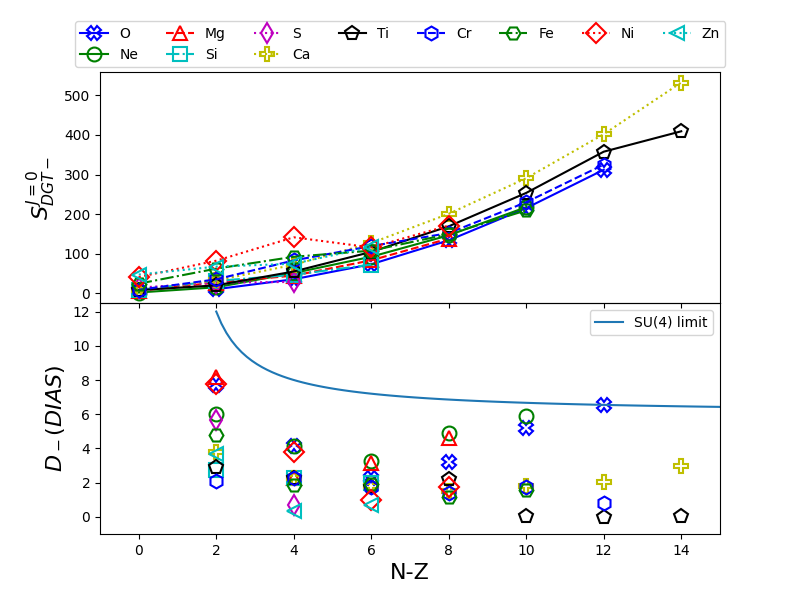}
	\caption{$D_-(DIAS)$ in comparison with $S^{J=0}_{DGT-}$, for even-even nuclei in $1s0d$ and $1p0f$ shells. Scattered data points are for $D_-(DIAS)$, while those linked with lines are for $S_{DGT-}$ in the $J=0$ channel. }
	\label{fig:DIAS}
\end{figure}

In Fig. \ref{fig:DIAS} we present theoretical results of DGT strengths on the DIAS, for even-even $N\geq Z$ isotopes of O, Ne, Mg, Si, S, Ca, Ti, Cr, Fe, Ni, Zn.
While $S^{J=0}_{DGT-}$ increases in a parabolic manner following the model-independent term in Eq.~(\ref{eqn:DGTSR-J0}), $D_-(DIAS)$ appears irrelevant to $N-Z$.
As a result, the fraction of $D_-(DIAS)$ in the total strength function becomes negligible when $N-Z \geq 4$.
That is predicted by the Wigner SU(4) in Eq. (\ref{eqn:SU4-DGT-DIAS},\ref{eqn:SU4-DGT-DGTS0}), i.e. $D_-(DIAS) \in (6,12]$, while $S^{J=0}_{DGT-}$ increases as $2(N-Z)(N-Z+1)$.

When $N-Z=0$, $D_-(DIAS)$ is blocked since the the initial nucleus has isospin $T_i = 0$.
When $N-Z=2$, SU(4) predicts that all $S^{J=0}_{DGT-}=D_-(DGT) = 12$, i.e. all DGT strengths are on the DIAS. 

\renewcommand{\arraystretch}{1.3}
\begin{table} [H]
	\centering
	\caption{ The fraction of DIAS strength in $S^{J=0}_{DGT-}$, for $N-Z=2$ nuclei. }
	\begin{tabular}{ccccccc}
		\hline\hline 
                           & $^{18}$O & $^{22}$Ne & $^{26}$Mg & $^{30}$Si & $^{34}$S & $^{38}$Ar	 \\
	
$D_-(DIAS)/S^{J=0}_{DGT-}$ & $74.1\%$ & $39.3\%$ & $29.9\%$  & $8.8\%$ & $2.1\%$ & $18.4\%$  \\
\hline
                           & $^{42}$Ca & $^{46}$Ti  & $^{50}$Cr & $^{54}$Fe & $^{58}$Ni & $^{62}$Zn \\
$D_-(DIAS)/S^{J=0}_{DGT-}$ & $44.7\%$  & $14.5\%$  & $5.9\%$  & $7.7\%$ & $9.4\%$ & $5.4\%$ \\
		\hline\hline
	\end{tabular}

	\label{tab-DIAS}
\end{table}

In Tab. \ref{tab-DIAS}, we present the fraction of DIAS contribution in $S^{J=0}_{DGT-}$, for $N-Z=2$ nuclei.
Three features can be identified: (i) as the valence particles increase, the DIAS fraction decreases; (ii) in the higher $1p0f$ shell, the DIAS fraction is not as significant as in the lower $1s0d$ shell;(iii) the fraction can be enhanced when approaching closed shells as in $^{38}$Ar, or the $0f_{7/2}$ subshell as in $^{58}$Ni.
The dominance of the DIAS contribution in $^{18}$O, $^{22}$Ne, $^{26}$Mg, $^{42}$Ca and $^{46}$Ti may results in a narrow strong peak in experiments, which is reminiscent of the ``low-energy super GT state" \cite{Fujita2014,PhysRevC.91.064316}, which was also observed in particle-type $N-Z=2$ nuclei, but not in hole-type $N-Z=2$ nuclei.
It was argued that $T=0$ interactions can pull the GT strengths to a lower excitation energy\cite{Bai2013,Fujita2014}, bending the strength function toward the SU(4) predictions.
In our context, it is possible that the ground state of particle-type $N-Z=2$ nuclei tends closer to an SU(4) weight state, as a result its DIAS(or its mirror state) is a ``super DGT state".

\section{Summary and Outlook} \label{sec-summary}
In summary, we present a systematic investigation on the double Gamow-Teller sum rules. Based on known DGTSR formulas, we suggest a model-independent rule, Eq.~(\ref{eqn:3GTSR2}), among three sum rules $S_{GT-}, S^{J=0,2}_{DGT}$, which can be used by experimentalists directly.
If experimental data of these three sum rules can be obtained in the future, comparison between GT quenching and DGT quenching can be realized.

Then we computed total strengths of double Gamow-Teller transition, i.e. $S_{DGT\pm}$ in the method of projection after variation of pair condensates (PVPC).
Benchmarking against shell-model results available in the literature shows that $S_{DGT\pm}$ from PVPC are accurate for semi-magic nuclei.
For open-shell nuclei without systematic shell-model results in the literature, we present DGTSR data in the $1s0d$ and $1p0f$ major shells.
The results show that, when $N-Z \geq 8$, $S_{DGT+}$ is below 4\% of $S_{DGT-}$, and the model independent term in Eq.~(\ref{eqn:DGTSR-J0},\ref{eqn:DGTSR-J2}) exhausts more than $85\%$ of $S^{J=0}_{DGT-}$, and more than $66\%$ of $S^{J=2}_{DGT-}$ respectively.

After elimination of error caused by PVPC's overestimating $S_{GT+}$ values, $S'_{DGT}$ values approach the MIT faster.
In future work, the Generator Coordinate Method, or the variation after projection techniques, can be used to improve the accuracy of PVPC states, and promote the quality of $S_{DGT}$ data.

The strength on the DIAS final state, i.e. $D_-(DIAS)$, remains on the magnitude $1\sim10$, irrelevant to $N-Z$, while the total strength $S_{DGT-}$ exhibits a parabolic growth as $N-Z$ increases.
In $N-Z=2$ nuclei with a few valence particles, $^{18}$O, $^{22}$Ne, $^{26}$Mg, $^{42}$Ca, $^{46}$Ti, $D_-(DIAS)$ dominates the strength function, suggesting possible ``super DGT states" similar to ``super GT states"~\cite{Fujita2014,PhysRevC.91.064316}.
These features are qualitatively in alignment with the predictions of the SU(4) symmetry.
%In future work, by GCM with pair condensates, or variation after projection techniques, the quality of PVPC ground states can be further improved, therefore the DGT sum rules can be more accurate.

%%%%%%%%%%%%%%%%%%%%%%%%%%%%%%%%%%%%%%%%%%
\authorcontributions{Conceptualization, Yi Lu and Calvin W. Johnson; methodology, Yi Lu and Yang Lei; formal analysis, Hong-Jin Xie; investigation, Shu-Yuan Liang; data curation, Hong-Jin Xie and Shu-Yuan Liang; writing---original draft preparation, Hong-Jin Xie; writing---review and editing, Yi Lu and Calvin W. Johnson; project administration, Yi Lu. All authors have read and agreed to the published version of the manuscript.}

\funding{Yi Lu acknowledges support by the Natural Science Foundation of Shandong Province, China (ZR2022MA050), the National Natural Science Foundation of China (11705100, 12175115).
Yang Lei is grateful for the financial support of the Sichuan Science and Technology Program (Grant No. 2019JDRC0017), and the Doctoral Program of Southwest University of Science and Technology (Grant No. 18zx7147). 
This material is also based upon work (Calvin W. Johnson) supported by the U.S. Department of Energy, Office of Science, Office of Nuclear Physics, under Award Number  DE-FG02-03ER41272.}

\dataavailability{
The code {\it PandasCommute} for one-body transition sum rule operator can be found at https://github.com/luyi07/PandasCommute.git.
The shell model code {\it BIGSTICK} can be found at https://github.com/cwjsdsu/BigStickPublick.git. 
The pair condensate code {\it PVPC} is not ready to be published yet, but the sum rule data, and documented shell scripts for {\it BIGSTICK} and {\it PandasCommute} can be found in the supplementary files.
} 

\conflictsofinterest{The authors declare no conflicts of interest.} 

%%%%%%%%%%%%%%%%%%%%%%%%%%%%%%%%%%%%%%%%%%
%% Optional

%% Only for journal Encyclopedia
%\entrylink{The Link to this entry published on the encyclopedia platform.}

\abbreviations{Abbreviations}{
The following abbreviations are used in this manuscript:
\\

\noindent 
\begin{tabular}{@{}ll}
DGT & Double Gamow-Teller\\
DGTSR & Double Gamow-Teller Sum Rule\\
DIAS & Double Isospin Analogue States \\
DCX & Double Charge Exchange
\end{tabular}
}

%%%%%%%%%%%%%%%%%%%%%%%%%%%%%%%%%%%%%%%%%%
%\isPreprints{}{% This command is only used for ``preprints''.
\begin{adjustwidth}{-\extralength}{0cm}

\reftitle{References}

% Please provide the correct journal abbreviation (e.g. according to the “List of Title Word Abbreviations” http://www.issn.org/services/online-services/access-to-the-ltwa/).

\bibliography{ref-papers}
%\bibliographystyle{mdpi}

%=====================================
% References, variant B: internal bibliography
%=====================================

% If authors have biography, please use the format below
%\section*{Short Biography of Authors}
%\bio
%{\raisebox{-0.35cm}{\includegraphics[width=3.5cm,height=5.3cm,clip,keepaspectratio]{Definitions/author1.pdf}}}
%{\textbf{Firstname Lastname} Biography of first author}
%
%\bio
%{\raisebox{-0.35cm}{\includegraphics[width=3.5cm,height=5.3cm,clip,keepaspectratio]{Definitions/author2.jpg}}}
%{\textbf{Firstname Lastname} Biography of second author}

% For the MDPI journals use author-date citation, please follow the formatting guidelines on http://www.mdpi.com/authors/references
% To cite two works by the same author: \citeauthor{ref-journal-1a} (\citeyear{ref-journal-1a}, \citeyear{ref-journal-1b}). This produces: Whittaker (1967, 1975)
% To cite two works by the same author with specific pages: \citeauthor{ref-journal-3a} (\citeyear{ref-journal-3a}, p. 328; \citeyear{ref-journal-3b}, p.475). This produces: Wong (1999, p. 328; 2000, p. 475)

%%%%%%%%%%%%%%%%%%%%%%%%%%%%%%%%%%%%%%%%%%
%% for journal Sci
%\reviewreports{\\
%Reviewer 1 comments and authors’ response\\
%Reviewer 2 comments and authors’ response\\
%Reviewer 3 comments and authors’ response
%}
%%%%%%%%%%%%%%%%%%%%%%%%%%%%%%%%%%%%%%%%%%
\PublishersNote{}
%\isPreprints{}{% This command is only used for ``preprints''.
\end{adjustwidth}
\end{document}